# Hounsfield-based Automatic Evaluation of Volumetric Breast Density on Radiotherapy CT-Scans

Deborah E. M. Akuoko, Dr. Eliana Vasquez Osorio, Prof. Marcel Van Herk, Dr. Marianne Aznar

*Abstract*—Radiotherapy is an integral part of treatment for many patients with breast cancer. However, side effects can occur, e.g., fibrosis or erythema. If patients at higher risks of radiation-induced side effects could be identified before treatment, they could be given more individual information about the risks and benefits of radiotherapy. We hypothesize that breast density is correlated with the risk of side effects and present a novel method for automatic evaluation based on radiotherapy planning CT scans. Methods: 799 supine CT scans of breast radiotherapy patients were available from the REQUITE dataset (www.requite.eu). The methodology was first established in a subset of 114 patients (cohort 1) before being applied to the whole dataset (cohort 2). All patients were scanned in the supine position, with arms up, and the treated breast (ipsilateral) was identified. Manual experts contour available in 96 patients for both the ipsilateral and contralateral breast in cohort 1. Breast tissue was segmented using atlas-based automatic contouring software, ADMIRE® v3.4 (Elekta AB, Sweden). Once validated, the automatic segmentation method was applied to cohort 2. Breast density was then investigated by thresholding voxels within the contours, using Otsu threshold and pixel intensity ranges based on Hounsfield units (-200 to -100 for fatty tissue, and -99 to +100 for fibro-glandular tissue). Volumetric breast density (VBD) was defined as volume of fibro-glandular tissue / (volume of fibro-glandular tissue + volume of fatty tissue). A sensitivity analysis was performed to verify whether calculated VBD was affected by choice of breast contour. In addition, we investigated correlation between volumetric breast density (VBD) and patient age and breast size. VBD values were compared between ipsilateral and contralateral breast contours. Results: Estimated VBD values were 0.40 (range 0.17-0.91) in cohort 1, and 0.43 (0.096-0.99) in cohort 2. We observed ipsilateral breasts to be denser than contralateral breasts. Breast density was negatively associated with breast volume (Spearman: R=-0.5, p-value < 2.2e-16) and age (Spearman: R=-0.24, p-value = 4.6e-10). Conclusion: VBD estimates could be obtained automatically on a large CT dataset. Breast density may not entirely be explained from variables such as patients' age or breast volume. Future work will focus on assessing the usefulness of VBD as a predictive variable for radiation-induced side effects.

*Keywords*—Breast cancer, automatic image segmentation, radiotherapy, big data, breast density, medical imaging.

## I. INTRODUCTION

It is a widely accepted notion that breast cancer risks can be predicted by breast density. Recent studies found that women with a mammographic density percentage above 50% tend to have a higher breast cancer risk which could be 1.8 to 6.0 times that of women with very lower breast density [1]. Also changes in breast density over time have been associated with changes in cancer risk [2]–[5]. Some studies that suggest that dietary factors, hormone exposure and reproductive factors are associated with breast density [6]-[11]. Other research reveals an association between higher mammography density and longer use of hormone therapy [12]. The breast is made of three main constituents: skin, fat and fibro glandular tissue [13]. Research comparing breast density reading from CT scans to mammograms for same patient revealed consistency between densities from the two imaging modalities. It was also suggested from preliminary results that upon further validation, breast density calculated from CTs could provide additional information about breast cancer risks [14], [15]. Other studies have revealed an increase in BI-RADS density category over a three-year period is associated with an increase in breast cancer risk and the vice versa [4].

Breast density has been proven in several studies as a key risk factor for breast cancer development [2], [16]-[18]. An important question of how best to determine which breast composition is dense remains unanswered. Current qualitative means of breast density measurements include the BI-RADS category. This mammographic density categories, based on considerable inter-observer variability categorizes density as percentages in its 4th edition [19]-[21]. In a study conducted to evaluate volumetric breast density correlation to qualitative BI-RADS categories, a good correlation was found between the two methods where the BI-RADS density categories were assigned by two radiologists [22]. It was therefore concluded that fully automated volumetric method may be used to quantify breast density on routine mammography. However, in another study [23], the correlation between BI-RADS scores from European radiologists and volumetric density only showed a moderate agreement between the two metrics of breast density assessment. Thus, though volumetric breast densities have proven to be useful, further investigation could help to fully understand how they may differ in breast density interpretation compared to the BI-RADS category [19]-[21]. Some unestablished hypotheses about breast density include its relationship to the malignancy of tumours [24]-[27], the likelihood of a patient redoing their surgery [28], [29] and its probable relationship to age. or mean glandular dose estimation [30]-[36]. In this project we will explore the correlation between breast density and variables such as patient age and volume of breast. Despite previous studies that have investigated breast composition, most of the work done used qualitative evaluation in investigating composition of fatty tissue and fibro glandular tissue [20], [37], and we aim to evaluate breast density quantitively.

## II. MATERIALS AND METHODS

*Dataset*—Radiotherapy planning CT scans from the REQUITE study were acquired for breast cancer patients treated in prone and supine positions. Data for all received scans were anonymised. We first validated our method using 114 CTs and used 947 CTs for analysis. Some CTs had missing contours which we resolved with auto-contouring.

*Processing Data for Breast Delineation Study*— 18/114 patients in cohort 1 were scanned in prone position. 55 of the 96 patients treated in supine position were treated for right sided breast cancer and the other 41 of the 96 patients were treated for left sided breast cancer

REQUITE data had a variety of standard and practices from several institutions. E.g., some institutions do not contour CTV routinely. Instead, surrogate PTV volumes are used. It was important for us to ensure that all the breast volumes were contoured in a consistent manner, however it was not practical to contour so many patients manually. Thus, we investigated and adopted an atlas based automatic contouring approach (research version of advanced medical image registration engine, ADMIRE® prototype by Elekta). Quality assessment of CTVs used as training data for atlas contouring was by visual evaluation checks with ESTRO contouring guidelines.

From each treatment position sample, 10 contoured CTs were chosen to create an atlas to represent ground truth. CTs with actual CTV contours were not automatically recontoured. Software for processing CT images was Worldmatch, our in-house medical image processing application. ROI of processed images were burned to Nifty files for further analyses of CTs intensity distribution in python, using SimpleITK.

*Image Similarity*— ADMIRE® contours were compared to manual contours by clinicians in a validation dataset of cohort 1 scans.

*Processing Data for Breast Density Study*— A threshold algorithm was used to visualize the composition of fatty tissue and fibro glandular tissue based on pixel intensities (PI) greater than -200 and less than +100 to eliminate abnormally higher bone and/or muscle intensities (Fig.1). The following Hounsfield units (HU) were used to calculate fatty and fibro glandular volumes:
HU: -210 to -100 for fatty tissue,
HU: -99 to +100 for fibro- glandular tissue.
Volumetric breast density (VBD) was defined as volume of fibro-glandular tissue / (volume of fibro-glandular tissue + volume of fatty tissue).

*Breast density study*—Ipsilateral breast CTs had been operated with possible fibrosis or edema present. Contralateral breast densities were therefore compared to ipsilateral.

## III. RESULTS

We recorded an average DSC of 0.89 ADMIRE® in comparison manual contouring, and we recorded strong similarities between medians of recorded volumes from ADMIRE® and manual delineations

*Thresholding Fatty tissue and Fibro glandular tissue*— analyses of breast density study showed an average of 56.0% fatty tissue in ipsilateral breast and 44.0% fibro glandular tissue with no outliers seen when pixel intensity was visualised (Fig.1). These were like data from general population [37] (*12% fatty, 45% scattered areas of density, 32% heterogeneously dense and 11% extremely dense*) [37]

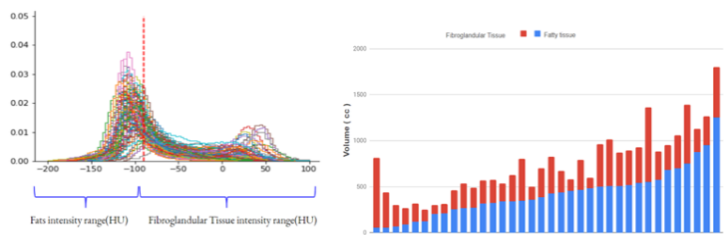

Fig. 1 Left: distribution of pixel intensity in breast contours. Right: Fatty tissue to fibro glandular tissue per breast contour on cohort I.

*Structural composition of breast*— We delved a bit deeper into what the breast contours were really made of aside from fatty tissue and fibro glandular tissue. We calculated breast volume in two forms. First as a whole breast volume (included fatty tissue, fibro glandular tissue and other connective tissues within) and as a GF breast Volume (Included only fibro-glandular tissue and fatty tissue). We further investigated if breasts composed only of fats and fibro glandular tissue? Our first step was a visual inspection (Fig.2). It was visually seen that the breast was mainly composed of fibro glandular and fatty tissue (Fig.2). However, there was indication of small amounts of other tissues present. We compared medians of the two breast volumes (WSR, Wilcoxon signed rank test, p-value=0.85). Not much difference seen in visual comparison of volumes in a chart (Fig.3).

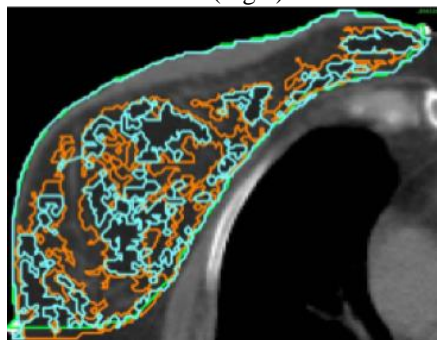

Fig. 2 Contours of Fibro glandular tissue and fat tissue within sample breast from cohort 1. (Fatty tissue: orange, fibro-glandular tissue: turquoise, Ipsilateral breast contour: green).

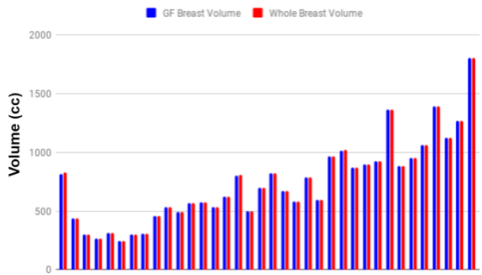

Fig. 3 Whole versus GF breast volumes.

*Validation of VBD for ipsilateral and contralateral breast contours—* we next compared Ipsilateral and Contralateral breast densities. We found evidence that medians of two sets of volumes are not equal (WSR, p-value = 1.449e-07), with higher ipsilateral breast volumes.

Possible explanation for this was assumed to be fibrosis. This assumption was made because most breast cancer patients who are yet to receive radiotherapy would typically have had surgery, which can result in fibrosis. (The presence of fibrosis in a patient breast prior to the REQUITE study was neither specified in the inclusion or exclusion criteria.)

*Breast Density Study Results and Analyses—* This section will include analyses of results for breast densities calculation in cohort 2. We used ADMIRE® to automate propagation of CTV contours on Supine patients using atlases created from our validation method. It was found that all prone patients had actual CTV delineations, thus we used manual contours in all prone scans to calculate breast density. Table I shows a descriptive summary of key variables that would be used in our analyses. A series of questions were asked to acquire three main sets of results. First, we investigated relationship between patient age and breast density (Fig.4, 5). We used Spearman correlation and regression to assess correlation between the two variables.

TABLE I
SUMMARY OF SIGNIFICANT VARIABLES FROM COHORT 2 SUPINE CTs

| | MIN. | 1ST QU. | MEDIAN | MEAN | 3RD QU. | MAX. |
|---|---|---|---|---|---|---|
| AGE | 23.00 | 50.00 | 59.00 | 58.45 | 66.00 | 90.00 |
| IPSILATERAL BREAST DENSITY | 0.096 | 0.32 | 0.43 | 0.46 | 0.57 | 0.99 |
| IPSILATERAL BREAST VOL. | 23.30 | 431.30 | 641.40 | 709.30 | 901.80 | 2449.50 |
| IPSILATERAL FATTY TISSUE VOL. | 1.85 | 184.81 | 347.85 | 394.84 | 538.89 | 1573.97 |
| CONTRALATERAL BREAST DENSITY | 0.085 | 0.25 | 0.35 | 0.40 | 0.49 | 0.99 |
| CONTRALATERAL BREAST VOLUME | 30.84 | 430.58 | 650.73 | 701.92 | 882.06 | 2419.22 |

*Correlation assessment between patients' age and VBD—*Higher breast density is often associated with younger women and other hereditary factors or menopausal status. In a study to examine mammographic breast density and patient age [38], an inverse relationship was found between patients' age and their breast density. The study also reported outliers at the age extremes. To further understand the extent to which patient VBD can be explained by their age, we used a spearman rank correlation (Fig.4) to analyse the two variables (SRC, P-value=4.2e-10).

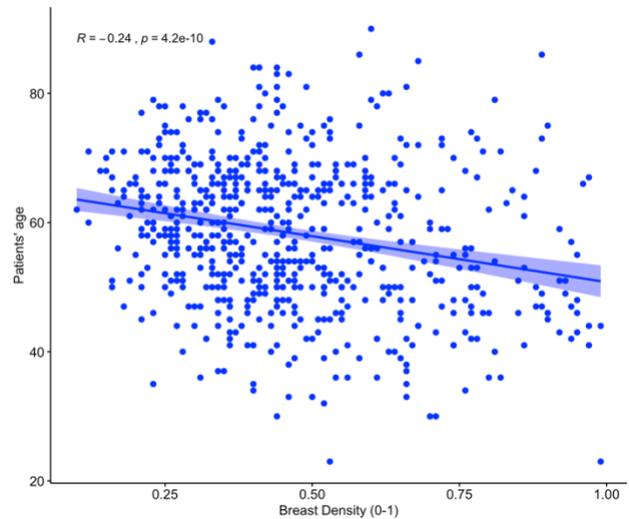

Fig. 4 SRC between patient age and their corresponding breast density.

*Correlation assessment between patients' age and VBD (for volumes >500cc)—*To understand correlation outcomes for the population with normal breast volumes, we repeated the test, excluding very low breast volumes which we may not be certain of its causes (See section IV - limitations). This returned an even weaker correlation, R=-0.1 (Fig.5) and p-value 0.029. Both correlation tests returned weak coefficients (Rho == -0.24 and -0.1). We had evidence to claim that patients' ages may not be the best way to explain their breast density.

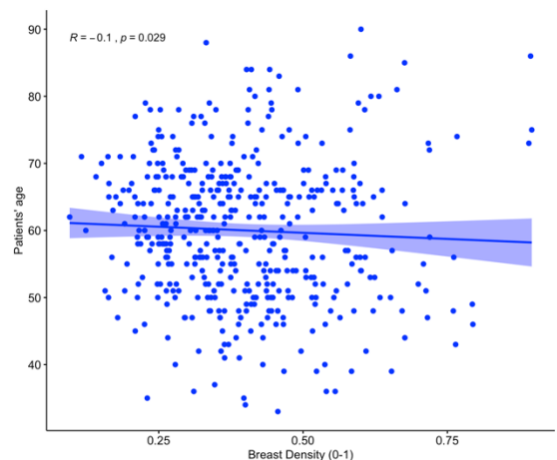

Fig. 5 SRC between patient age and corresponding breast density (for volumes >500cc)

*Ipsilateral and contralateral breast volumes*— There was no significant difference seen (Fig.6) between volume of the contralateral (non-operated) breasts and the ipsilateral (operated) breast.

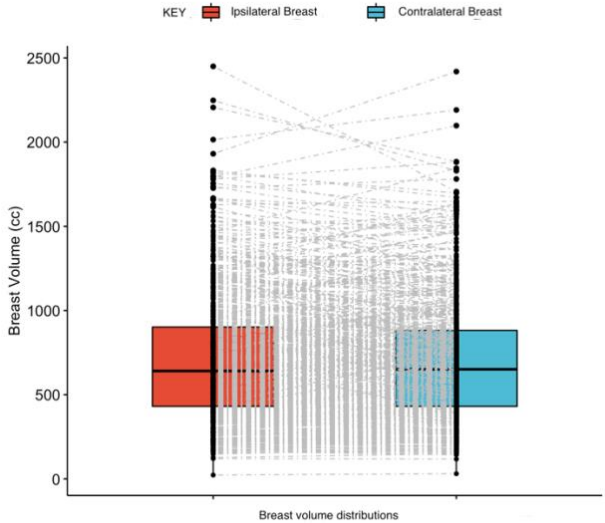

Fig. 6 Breast volumes (cc) distribution - ipsilateral vs contralateral delineations

*Test to compare ipsilateral and contralateral VBD*— We compared differences between ipsilateral and contralateral VBDs (Fig.7). Another test was performed to investigate difference in medians of the two VBD samples (WSR, P-value=2.2e-16).

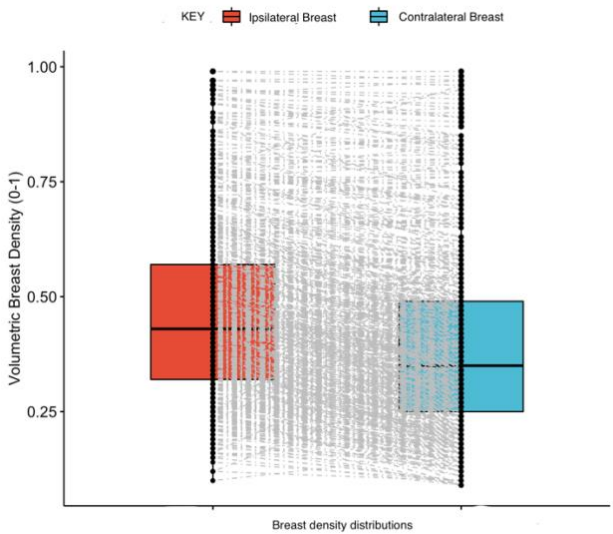

Fig. 7 Breast density (0-1) for ipsilateral and contralateral contours

*Correlation assessment between patients' breast volume and breast density*— Higher breast density is often associated with higher breast cancer risks. To further understand the extent to which VBD can be explained breast volume, performed another test (Fig.8, p-value<2.2e-16).

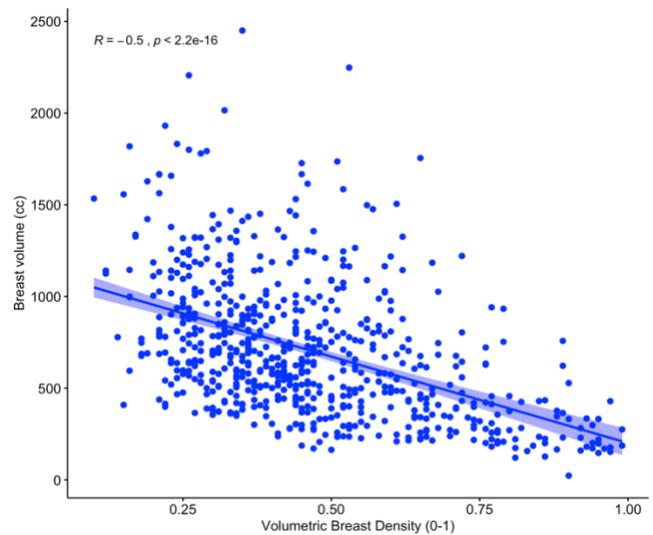

Fig. 8 SRC between breast volumes (cc) and corresponding breast density

*Correlation between patients' breast volume and breast density (for volumes >500cc)*— we repeated the test (Fig.9), excluding very low breast volumes which we may not be certain of its causes (See section IV - limitations). This returned an even weaker correlation coefficient, R=-0.19. Both tests returned relatively weak negative correlation coefficients (-0.5, -0.19), supporting our claim that patients' breast volumes may not be the best way to explain their breast density.

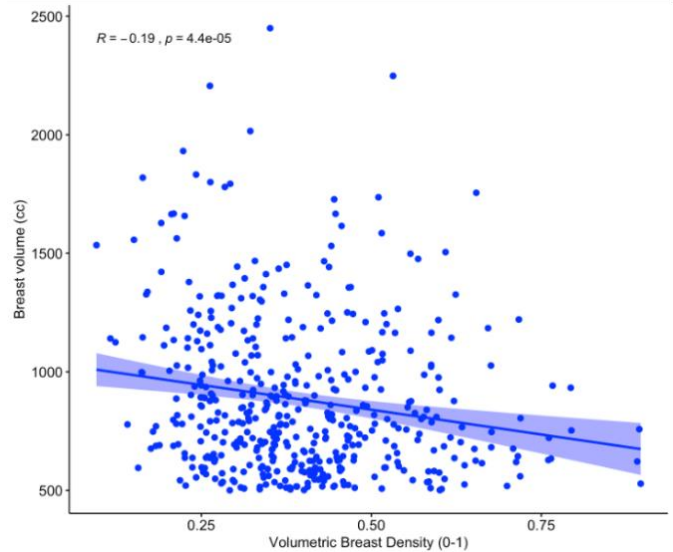

Fig. 9 SRC between breast volumes (cc) and corresponding breast densities (for volumes >500cc).

IV. DISCUSSION

*Imperfections*— we discuss incidents that were not necessarily limitations but could either have been much better or it required extra effort and considerations from us for it to become suitable for the overall output of the project

Inter-observer delineation variability— Pre-treatment radiotherapy CTs used in this project, had been manually delineated by clinical experts. Manual delineation is viewed as a gold standard in most research projects, yet it is not exempted from Inter Observer Variability (IOV) or uncertainties among clinical experts. Despite the availability of satisfactory documentation in radiation oncology [39], it is evident that delineation uncertainties can cause potential RT treatment errors in dose delivery [40] and even have possible implications for patients' toxicities [39] the main purpose of manual delineations in our work was to define a region of interest (ROI) from which we could calculate breast composition to investigate breast densities and how it correlates to patient outcomes. In the interest of IOV, we performed sensitivity tests in our work to verify that the density of the breast did not change due of the type of automatic contouring approach used. Also, since our work was mainly focused on breast composition rather than breast tumours or dose delivery, we had reasons to assume that the impact of IOV on dose delivery or tumour delineation accuracy would be small on our results. Thus, implications of delineation uncertainties for patients' toxicities [39] were not considered very significant to our breast density calculation outcome.

Possible treatment on both breasts— Our dataset consisted of breast cancer patients. It was assumed that as the patients were being prepared for RT treatment, they would have had at least some sort of surgery on the ipsilateral breast. This was why we considered delineating the contralateral breasts as well. But we were not 100% certain that the contralateral breast as at the time has not had a previous tumour or treatment. Though REQUITE had extensive documentation on the inclusion and exclusion criteria, our findings on unexpected data caused us to leave room for possibilities but also assumed that it should most likely have been excluded before we accessed the data.

Unexpected data found— We found some data that did not fit into the REQUITE data inclusion criteria. For instance, it was stated that women who had breast implants had been excluded but one CT was seen to have breast implants on visual inspection. This was excluded. Other unexpected data that was found and excluded by us include scans without Radiotherapy Treatment Plan (RTP), one patient that had a rotated treatment position, scans without any structure files and patients' scans with ROI delineated at unexpected positions. The presence of these unexpected data forced us to individually inspect all scans before batch processing.

Lack of naming standards for data— We found the implication of receiving data from multiple institutions to be quite mixed. Because it was an advantage that we had such large datasets from varying sources or health institutions. However, it was the most time-consuming task to clean this data. The reason was not just because different health institutions used different languages within their individual jurisdiction. There was no convention or standard of naming contours. For this reason, even after overcoming the language barrier of understanding what the names meant, we still could not use an easy automatic way to classify delineations of interest because there was no common ground for how to name delineation. We consider it a success that despite this challenge, we dedicated time and resources to clean the data to a level that made it suitable and accurate for our research.

Auto contouring Uncertainties— We have observed from our results sections that both ADMIRE® have satisfactorily high dice coefficients and very similar volume distributions to our ground truth. However, we are unable to rule out the fact that the algorithm could possibly encounter an error or delineate a CTs unsatisfactorily. Quality checks were conducted to help keep calculated volumes in check. Thus, impacts of this uncertainty on our results may not be very large.

Risk of mastectomy— As the patients used in this study were being prepared for radiotherapy, they most likely would have undergone surgery (A part of their breast would probably have been taken out). In our results, we observed patients with excessively small breast volumes who also happen to be extremely dense. The REQUITE data inclusion criteria stated mastectomy patients would have been excluded. But the observed unusually small breast volumes gives us reason to expect that there might be possibility of patients with mastectomy in our dataset.

*Successes—* Irrespective of some imperfect situations, our work achieved successes. The below sections will discuss how we succeeded in calculating breast composition and the potential findings we hope to uncover once we have access to the outcome data.

Volume and density measurement— Breast Density information is very key in development of risk stratification models for development of breast cancer. However, it is currently unknown how much impact breast density has on patient toxicity reactions post-radiotherapy (RT). With an aim to identify the correlation between post RT outcomes of patients with respect to density of their breasts, we started with an investigation of breast composition. To decide how to separate fibro-glandular tissue from fatty tissue, we reviewed the recent literature about the estimated distribution of breast composition in women receiving mammography for breast cancer screening. This guided us in choosing the appropriate HU intensity range. We assessed two atlas-based segmentation software to determine which of the two was most efficient especially for our automatic contour propagation work. We were able to automatically contour ipsilateral and contralateral breasts that were not available to aid our breast density investigations.

Impactful finding— From our breast density study, we confirmed that breast density decreased with age (which is well known from mammography studies), however, the relationship between age and density was very weak. Hence, density is best estimated from actual images of each woman's breasts and cannot be reliably estimated solely from clinical factors such as age. In future work, when building a risk model for radiation-toxicity, we suggest that CT based density should be investigated as an independent variable.

*Limitations—* We had a couple of limitations during this project. Most of which were in relation to the dataset or acquisition of some parts of the dataset which was needed. In the sub sections below, specific limitation has been discussed. How it imparted the project as well as how it could have been better is also discussed

Unavailable patient outcome and age data— Due to certain factors beyond the control of the project, data relating to toxicity outcomes which were recorded by the REQUITE study was not available at the time of analyses. The main impact this had on our research was inability to test our main hypothesis; 'Dense breasts could have a higher risk of post-radiotherapy toxicities' by the end of the project as mentioned in section 1.8.

Choice of HU threshold— To measure breast density, we first defined intensity ranges that were representative of the major composition of the breast (fibro-glandular tissue and fatty tissue). How we chose this has been explained in section 2.4, however, we have not been able to verify the reliability of this choice of intensity range in this work. A good way to validate this would be to acquire equivalent mammographic images for patients on our cohort. Purpose of this will be to calculate mammographic densities and compare them to densities of current CTs to measure how they relate. This limitation has been stated as possible future work.

Contralateral breast density validation— Due to availability of fewer contralateral breast delineations (section 2.4), our validation dataset for contralateral breasts was much smaller. Comparison of automatically contoured contralateral VBD to manually contoured contralateral VBD shows the two datasets do not vary as much. However, we unable to generalise this for larger population due to the significantly smaller number of samples used in validation.

*Reflections—* as a reflection of our methodology, we report on a couple of findings during the work.

Choice of Breast Density Calculation Approach— There are quite a few methods available for calculating breast density. We chose to use Volumetric Breast Density (VBD) approach for the following reasons. First, volumetric breast density from mammograms have potential to be used in objective risk models [42] and has good correlation with visual assessment for BI-RADS ($5^{th}$ and $4^{th}$ Editions) [43] In this project we calculated breast density from CTs and not mammograms. Previous work to compare volumetric densities to BI-RAD categories reveal a good correlation seen between volumetric density and BI-RADS density categories [22]. We, therefore, state in our future developments, the need to compare mammographic densities of patients with the calculated VBDs to validate how our breast densities conform to general BI-RAD standards [37]. The second reason we chose to use VBD is to avoid the possible impact of breast composition that was neither fatty nor fibro-glandular tissue. There were parts of the breast composition that formed a much smaller volume. Possible definition of this smaller volume could be connective tissue [41] In other breast density calculation methods that are based on pixels' intensity of ROI, we would calculate average of all intensities within whole breast as density. This would include intensities of this small non major breast composition. To avoid unknown or uncertain impacts of non-major composition tissue on our breast density results, we considered using VBD which only considers the volumes of major breast compositions.

Breast Volume vs Breast Density— Our data sample showed some evidence of a weak correlation between breast volume and breast density. Weaker correlation coefficients were recorded both for all volumes and vol > 500cc, breast volume of breast cancer patients may not be the best variable to explain their breast density.

V. FUTURE DEVELOPMENTS

One suggestion for future research is to acquire corresponding toxicity outcome data for the sample population to investigate the relationship between patients' breast density and their toxicity outcomes. Our finding could also enable us to develop a model for predicting patient toxicity based on their breast density. Clinical significance of this model includes helping oncologists to develop tailored treatments for patients who are to receive radiotherapy. Due to unavailability of corresponding mammograms for the breast CTs scans used in this project, we would recommend that patient mammograms would be acquired in future, to help verify breast densities recorded from our cohort with BI-RADS descriptors and to establish how recorded densities correspond to general standards.

VI. CONCLUSION

ADMIRE® was reliable in processing our data and we leverage more on it due to the automatic scripting feature. We were able to develop an approach to calculate breast densities from CT scans, which will be validated later (possibly with mammograms). We also found that breast density may not entirely be explained from variables such as patients' age or breast volume. As results, we can justify investigating our CT-estimated densities as variables in a risk prediction model

ACKNOWLEDGMENT

First, I would like to thank DARA Big data, led by Prof. Anna Scaife for the Newton Fund sponsorship opportunity to do this research. I am also most grateful to Dr. Alan McWilliam, who was not part of my supervisory team but still dedicated as much time and expertise to support my work.

REFERENCES

[1] Boyd, N., Dite, G., Stone, J., Gunasekara, A., English, D., McCredie, M., Giles, G., Tritchler, D., Chiarelli, A., Yaffe, M. and Hopper, J., 2002. Heritability of Mammographic Density, a Risk Factor for Breast Cancer. New England Journal of Medicine, 347(12), pp.886-894.
[2] Vachon, C., Pankratz, V., Scott, C., Maloney, S., Ghosh, K., Brandt, K., Milanese, T., Carston, M. and Sellers, T., 2007. Longitudinal Trends in Mammographic Percent Density and Breast Cancer Risk. Cancer Epidemiology Biomarkers & Prevention, 16(5), pp.921-928.
[3] Maskarinec, G., Pagano, I., Lurie, G. and Kolonel, L., 2006. A Longitudinal Investigation of Mammographic Density: The Multiethnic Cohort. Cancer Epidemiology Biomarkers & Prevention, 15(4), pp.732-739.

[4] Kerlikowske, K., Ichikawa, L., Miglioretti, D., Buist, D., Vacek, P., Smith-Bindman, R., Yankaskas, B., Carney, P. and Ballard-Barbash, R., 2007. Longitudinal Measurement of Clinical Mammographic Breast Density to Improve Estimation of Breast Cancer Risk. JNCI Journal of the National Cancer Institute, 99(5), pp.386-395.

[5] Guthrie, J., Milne, R., Hopper, J., Cawson, J., Dennerstein, L. and Burger, H., 2007. Mammographic densities during the menopausal transition. Menopause, 14(2), pp.208-215.

[6] El-Bastawissi AY, White E, Mandelson MT, Taplin SH. Reproductive and hormonal factors associated with mammographic breast density by age (United states). Cancer Causes Control 2000;11:955–63.

[7] Sterns EE, Zee B. Mammographic density changes in perimenopausal and postmenopausal women: is effect of hormone replacement therapy predictable? Breast Cancer Res Treat 2000;59:125–32.

[8] Van Gils CH, Hendriks JH, Otten JD, Holland R, Verbeek AL. Parity and mammographic breast density in relation to breast cancer risk: indication of interaction. Eur J Cancer Prev 2000; 9:105–11.

[9] Vachon CM, Kuni CC, Anderson K, Anderson VE, Sellers TA. Association of mammographically defined percent breast density with epidemiologic risk factors for breast cancer (United States). Cancer Causes Control 2000;11:653–62.

[10] Sala E, Warren R, Duffy S, Welch A, Luben R, Day N. High risk mammographic parenchymal patterns and diet: a case control study. Br J Cancer 2000;83:121–6.

[11] Greendale GA, Reboussin BA, Sie A, et al. Effects of estrogen and estrogen-progestin on mammographic parenchymal density. Postmenopausal Estrogen/Progestin Interventions (PEPI) Investigators. Ann Intern Med 1999;130:262–9

[12] Chen, F., Cheung, Y. and Soong, Y., 2010. Factors That Influence Changes in Mammographic Density With Postmenopausal Hormone Therapy. Taiwanese Journal of Obstetrics and Gynecology, 49(4), pp.413-418.

[13] Nelson, T., Cerviño, L., Boone, J. and Lindfors, K., 2008. Classification of breast computed tomography data. Medical Physics, 35(3), pp.1078-1086.

[14] l Salvatore, M., Margolies, L., Kale, M., Wisnivesky, J., Kotkin, S., Henschke, C. and Yankelevitz, D., 2014. Breast Density: Comparison of Chest CT with Mammography. Radiology, 270(1), pp.67-73

[15] Kerlikowske, K., Ichikawa, L., Miglioretti, D., Buist, D., Vacek, P., Smith-Bindman, R., Yankaskas, B., Carney, P. and Ballard-Barbash, R., 2007. Longitudinal Measurement of Clinical Mammographic Breast Density to Improve Estimation of Breast Cancer Risk. JNCI Journal of the National Cancer Institute, 99(5), pp.386-395.

[16] Boyd, N., Guo, H., Martin, L., Sun, L., Stone, J., Fishell, E., Jong, R., Hislop, G., Chiarelli, A., Minkin, S. and Yaffe, M., 2007. Mammographic density and the risk and detection of breast cancer. New England Journal of Mediccine, 356(3), pp.227-236.

[17] McCormack, V. and dos Santos Silva, I., 2006. Breast Density and Parenchymal Patterns as Markers of Breast Cancer Risk: A Meta-analysis. Cancer Epidemiology Biomarkers & Prevention, 15(6), pp.1159-1169.

[18] Yaffe, M., Boyd, N., Byng, J., Jong, R., Fishell, E., Lockwood, G., Little, L. and Tritchler, D., 1998. Breast cancer risk and measured mammographic density. European Journal of Cancer Prevention, 7, pp.S47-S56.

[19] Magny, S., Shikhman, R. and Keppke, A., 2019. Breast, Imaging, Reporting and Data System (BI RADS). StatPearls Publishing, Treasure Island (FL),.

[20] Ng, K. and Lau, S., 2015. Vision 20/20: Mammographic breast density and its clinical applications. Medical Physics, 42(12), pp.7059-7077.

[21] Spak, D., Plaxco, J., Santiago, L., Dryden, M. and Dogan, B., 2017. BI-RADS ® fifth edition: A summary of changes. Diagnostic and Interventional Imaging, 98(3), pp.179-190.

[22] Singh, T., Sharma, M., Singla, V. and Khandelwal, N., 2016. Breast Density Estimation with Fully Automated Volumetric Method:. Academic Radiology, 23(1), pp.78-83.

[23] Sartor, H., Lång, K., Rosso, A., Borgquist, S., Zackrisson, S. and Timberg, P., 2016. Measuring mammographic density: comparing a fully automated volumetric assessment versus European radiologists' qualitative classification. European Radiology, 26(12), pp.4354-4360.

[24] Bertrand, K., Tamimi, R., Scott, C., Jensen, M., Pankratz, V., Visscher, D., Norman, A., Couch, F., Shepherd, J., Fan, B., Chen, Y., Ma, L., Beck, A., Cummings, S., Kerlikowske, K. and Vachon, C., 2013. Mammographic density and risk of breast cancer by age and tumor characteristics. Breast Cancer Research, [online] 15(6), p.R104. Available at: <https://www.ncbi.nlm.nih.gov/pubmed/24188089>.

[25] Bertrand, K., Scott, C., Tamimi, R., Jensen, M., Pankratz, V., Norman, A., Visscher, D., Couch, F., Shepherd, J., Chen, Y., Fan, B., Wu, F., Ma, L., Beck, A., Cummings, S., Kerlikowske, K. and Vachon, C., 2015. Dense and Nondense Mammographic Area and Risk of Breast Cancer by Age and Tumor Characteristics. Cancer Epidemiology Biomarkers & Prevention, 24(5), pp.798-809.

[26] Boyd, N., Huszti, E., Melnichouk, O., Martin, L., Hislop, G., Chiarelli, A., Yaffe, M. and Minkin, S., 2014. Mammographic features associated with interval breast cancers in screening programs. Breast Cancer Research, 16(4).

[27] Yaghjyan, L., Colditz, G., Collins, L., Schnitt, S., Rosner, B., Vachon, C. and Tamimi, R., 2011. Mammographic Breast Density and Subsequent Risk of Breast Cancer in Postmenopausal Women According to Tumor Characteristics. JNCI Journal of the National Cancer Institute, 103(15), pp.1179-1189.

[28] Bani, M., Lux, M., Heusinger, K., Wenkel, E., Magener, A., Schulz-Wendtland, R., Beckmann, M. and Fasching, P., 2009. Factors correlating with reexcision after breast-conserving therapy. European Journal of Surgical Oncology (EJSO), 35(1), pp.32-37.

[29] Dieterich, M., Dieterich, H., Moch, H. and Rosso, C., 2012. Re-excision Rates and Local Recurrence in Breast Cancer Patients Undergoing Breast Conserving Therapy. Geburtshilfe und Frauenheilkunde, 72(11), pp.1018-1023.

[30] Gennaro, G., Sechopoulos, I., Gallo, L., Rossetti, V. and Highnam, R., 2015. Impact Of Objective Volumetric Breast Density Estimates On Mean Glandular Dose Calculations In Digital Mammography. ECR 2015 / C-1576. European Society of Radiology (ESR).

[31] Dance, D., Skinner, C., Young, K., Beckett, J. and Kotre, C., 2000. Additional factors for the estimation of mean glandular breast dose using the UK mammography dosimetry protocol. Physics in Medicine and Biology, 45(11), pp.3225-3240.

[32] Boone, J., 2002. Normalized glandular dose (DgN) coefficients for arbitrary x-ray spectra in mammography: Computer-fit values of Monte Carlo derived data. Medical Physics, 29(5), pp.869-875.

[33] Dance, D., Young, K. and van Engen, R., 2009. Further factors for the estimation of mean glandular dose using the United Kingdom, European and IAEA breast dosimetry protocols. Physics in Medicine and Biology, 54(14), pp.4361-4372.

[34] Dance, D., Young, K. and van Engen, R., 2010. Estimation of mean glandular dose for breast tomosynthesis: factors for use with the UK, European and IAEA breast dosimetry protocols. Physics in Medicine and Biology, 56(2), pp.453-471.

[35] Dance, D. and Young, K., 2014. Estimation of mean glandular dose for contrast enhanced digital mammography: factors for use with the UK, European and IAEA breast dosimetry protocols. Physics in Medicine and Biology, 59(9), pp.2127-2137.

[36] Sobol, W. and Wu, X., 1997. Parametrization of mammography normalized average glandular dose tables. Medical Physics, 24(4), pp.547-554.

[37] Sickles, EA, D'Orsi CJ, Bassett LW, et al. ACR BI-RADS® Mammography. In: ACR BI-RADS® Atlas, Breast Imaging Reporting and Data System. Reston, VA, American College of Radiology; 2013

[38] Adel, M., Rasigni, M., Bourennane, S. and Juhan, V., 2007. Statistical Segmentation of Regions of Interest on a Mammographic Image. *EURASIP Journal on Advances in Signal Processing*, 2007(1).

[39] Weiss, E. and Hess, C., 2003. The Impact of Gross Tumor Volume (GTV) and Clinical Target Volume (CTV) Definition on the Total Accuracy in Radiotherapy. Strahlentherapie und Onkologie, 179(1), pp.21-30.

[40] Vorwerk, H., Zink, K., Schiller, R., Budach, V., Böhmer, D., Kampfer, S., Popp, W., Sack, H. and Engenhart-Cabillic, R., 2014. Protection of quality and innovation in radiation oncology: The prospective multicenter trial the German Society of Radiation Oncology (DEGRO-QUIRO study). Strahlentherapie und Onkologie, 190(5), pp.433-443.

[41] Richard Hammerstein, G., Miller, D., White, D., Ellen Masterson, M., Woodard, H. and Laughlin, J., 1979. Absorbed Radiation Dose in Mammography. Radiology, 130(2), pp.485-491.

[42] Gubern-Mérida, A., Kallenberg, M., Platel, B., Mann, R., Martí, R. and Karssemeijer, N., 2014. Volumetric Breast Density Estimation from Full-Field Digital Mammograms: A Validation Study. PLoS ONE, 9(1), p.e85952.

[43] Youk, J., Kim, S., Son, E., Gweon, H. and Kim, J., 2017. Comparison of Visual Assessment of Breast Density in BI-RADS 4th and 5th Editions With Automated Volumetric Measurement. American Journal of Roentgenology, 209(3), pp.703-708.